\documentclass{iaus}

\usepackage{graphicx}
\usepackage{booktabs}

\title[Searching for binary central stars of planetary nebulae with Kepler]{Searching for binary  central stars of planetary nebulae with Kepler}

\author[D. Douchin et al.]{D. Douchin$^{1,2,3}$, G. H. Jacoby$^4$, O. De Marco$^{1,2}$, S. B.~Howell$^5$, M.~Kronberger$^6$}

\affiliation{$^1$Macquarie University Research Centre in Astronomy, Astrophysics \& Astrophotonics $^2$Department of Physics \& Astronomy, Macquarie University Australia $^3$GRAAL, Universit\'e Montpellier 2, France $^4$Giant Magellan Telescope,USA $^5$NASA Ames Research Center, USA $^6$Deep Sky Hunters Collaboration

}

\pubyear{2011}
\volume{283}  
\pagerange{119--126}
\jname{Planetary Nebulae: an Eye to the Future}
\editors{Arturo Manchado, Letizia Stanghellini \& Detlef Sch\"onberner, eds.}
\begin{document}

\maketitle


\begin{abstract}

The Kepler Observatory offers unprecedented photometric precision ( $<$1 mmag) and cadence for monitoring the central stars of planetary nebulae, allowing the detection of tiny periodic light curve variations, a possible signature of binarity. With this precision free from the observational gaps dictated by weather and lunar cycles, we are able to detect companions at much larger separations and with much smaller radii than ever before. We have been awarded observing time to obtain light-curves of the central stars of the six confirmed and possible planetary nebulae in the Kepler field, including the newly discovered object Kn 61, at cadences of both 30 min and 1 min. Of these six objects, we  could confirm for three a periodic variability consistent with binarity. Two others are variables, but the initial data set presents only weak periodicities. For the central star of Kn 61, Kepler data will be available in the near future.

\end{abstract}

\firstsection 

\section{Introduction}
In order to understand to what extent non-spherical planetary nebulae (PNe) are shaped by binary interactions, a fundamental test is the determination of the PN binary fraction. This is a notoriously difficult quantity to determine.
The short period binary fraction is estimated to be 15-20\% (\cite{Bond2000}, \cite{Miszalski2009}). This is biased to binaries with periods shorter than about $\sim 3$ days. 

The field of view of of the Kepler satellite of 14~x~14 square degrees, 4 PNe are known so far and 2 additional possible objects. We seek to detect periodic light variability due to binarity (irradiation effects, eclipses or ellipsoidal variablity) with an amplitude smaller than detectable by  ground observations in order to quantify the extent of the bias. If the currently known short period binary fraction of 15-20\% is accurate, we should see 1-2 binary PN in our Kepler sample.

\section{The PNe in the Kepler field}

\begin{figure}[!hbt]
\centering
\includegraphics[scale=0.16]{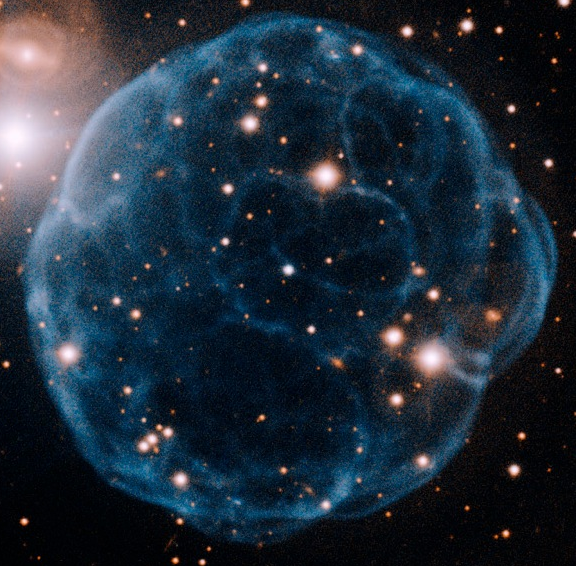}
\caption{Kn 61, newly detected PN in the Kepler field. Credit: Gemini Observatory and T.~A.~Rector (University of Alaska Anchorage). http://www.gemini.edu/node/11656} 
\label{kn61}
\end{figure}
There are four confirmed PNe in the Kepler field, including one (Pa 5) discovered by D.~Patchick in the framework of the Deep Sky Hunters (DSH) project (\cite{Jacoby2010}).
One additional PN candidate, Kn 61 was discovered in the Kepler field by M. Kronberger using DSS false color images also in the context of the DSH search (\cite{KronbergerIAU}), and then imaged by us in March 2011 with  a 10 min exposure at the 2.1m NOAO Kitt Peak observatory.  A Gemini Observatory image of the object is presented in Fig.~\ref{kn61}. From its structure, that is similar to PN A 43, it is classified as a probable PN, although a spectrum of the nebula is required before considering Kn 61 as a confirmed PN. The sixth object, PaTe 1, is classified as possible PN due to its unusual morphology.

\begin{table}[!htb]
\begin{center}
\begin{tabular}{l c c c c}
\toprule
Name        &     PN status     &  Period (days) &         Amplitude (mmag)      &             Notes              \\
\midrule
NGC 6826    &    confirmed      &     0.619     &            10                 &            Possible binary                                                     \\
NGC 6742    &    confirmed      &     0.37:     &            0.2:               &      Possible periodic variability                         \\
Pa 5        &    confirmed      &      1.12     &            4                  &            Possible binary, see \cite{Ostensen2010}                           \\
PaTe 1      &    possible       &      0.17     &            2                  &              Possible binary                                                   \\
A 61        & confirmed         &       1.47:     &   2:                         &          Variable                               \\
Kn 61       &   probable        &       -       &               -               &              Recently discovered, no data   \\
\bottomrule

\end{tabular}
\label{chart}
\caption{Preliminary results on the variability properties of the PNe in the Kepler field. The numbers followed by : are to be confirmed. Out of 5 possible or confirmed PNe, 4 show periodic light variations that could be attributed to binarity.}
\end{center}

\end{table}

 


\section{Early results and future work}

While the smallest detectable amplitude for flux variability is 0.1 mag for ground-based observations, Kepler can detect light variability of about 1 mmag. In Table 1 we show that  4 of the 5 confirmed and possible PNe that have already been observed by Kepler have periodic light variations consistent with binarity (though a proper analysis is required to confirm that the perodic variability of these objects is due to binarity) with amplitudes less than the ground-based detection limit. 

We will take thirty minute exposures of these six potential PNe in the Kepler field regularly throughout the year to detect potential long period flux variations. Wind and pulsations effects will be assessed with 1 minute sampling for each object over a single quarter. The advantages of this monitoring with respect to ground-based observations are a higher photometric precision (up to $1\mu$mag for bright stars and well monitored objects) and a longer time baseline with no day, bad weather nor lunar interruptions insuring a constant coverage and a better time sampling.

When the full data set is available, we will be able to impose some limits on how many short binaries may have avoided detection. Even with a small sample of PNe, we can put constraints on the actual short period binary fraction amongst the central stars of PNe.



\end{document}